\documentclass[prb,a4paper,superscriptaddress,twocolumn,showpacs]{revtex4}
\usepackage{graphicx}

\begin{document}

\title{Cooper-pair resonances and subgap Coulomb blockade
in a superconducting single-electron transistor}

\author{J.~J.~Toppari} 
\affiliation{Nanoscience Center, Department of Physics, University of Jyv\"askyl\"a,
P.O.~Box 35 (YN), FIN-40014 University of Jyv\"askyl\"a, Finland}

\author{T. K\"uhn}
\affiliation{Nanoscience Center, Department of Physics, University of Jyv\"askyl\"a,
P.O.~Box 35 (YN), FIN-40014 University of Jyv\"askyl\"a, Finland}

\author{A.~P.~Halvari}
\affiliation{Nanoscience Center, Department of Physics, University of Jyv\"askyl\"a,
P.O.~Box 35 (YN), FIN-40014 University of Jyv\"askyl\"a, Finland}

\author{J.~Kinnunen}
\affiliation{JILA and Department of Physics, University of Colorado, Boulder, 
Colorado 80309-00440, USA} 

\author{M.~Leskinen}
\affiliation{Nanoscience Center, Department of Physics, University of Jyv\"askyl\"a,
P.O.~Box 35 (YN), FIN-40014 University of Jyv\"askyl\"a, Finland}

\author{G.~S.~Paraoanu}
\affiliation{Nanoscience Center, Department of Physics, University of Jyv\"askyl\"a,
P.O.~Box 35 (YN), FIN-40014 University of Jyv\"askyl\"a, Finland}

\begin{abstract}

We have fabricated and measured superconducting single-electron transistors with Al leads and Nb 
islands. At bias voltages below the
gap of Nb we observe clear signatures of resonant tunneling of Cooper pairs, and of Coulomb blockade 
of the subgap currents due to linewidth broadening of the energy levels in the superconducting 
density of states of Nb. The experimental results are in good agreement with numerical simulations. 
\end{abstract}

\pacs{74.50.+r,73.23.Hk,73.40.Gk}

\maketitle

The single-electron transistor\cite{single} (SET) and its superconducting version is one of the most 
versatile tools in mesoscopic physics. It has been used for extremely sensitive charge 
measurements,\cite{rfset} for the construction of Cooper pair pumps and other adiabatic devices with 
applications in metrology,\cite{pumps} and more recently for building up superconducting quantum 
bits.\cite{chargequbits} 

The $IV$ characteristics of superconducting SETs present the usual features of quasiparticle 
tunneling (at voltages above $2\Delta_{\rm Nb}+2\Delta_{\rm Al}$), Josephson-quasiparticle tunneling
(at half of these values), and Josephson effect (around zero bias). These features have been 
thoroughly investigated by now by many groups and the physics of a charge transport at these bias 
voltages is well understood. However, at low bias voltages also other transport processes could 
become important and can alter the performance of Josephson-based devices. In this paper, we study 
two such processes appearing in our Nb-based SET: Resonant tunneling of Cooper pairs, and transport 
through states inside the gap of Nb (subgap currents).

We have fabricated Al/AlO$_x$/Nb/AlO$_x$/Al single electron transistors using a lithographic 
technique described elsewhere.\cite{us} Measurements were done using a small dilution refrigerator 
equipped with well-thermalized and electrically filtered measuring lines. The superconducting gaps 
obtained for Nb and Al ($\Delta_\mathrm{Nb}=1.4$ mV, $\Delta_\mathrm{Al}=0.2$ meV), and also the 
measured critical temperatures for Nb ($T_\mathrm{C,Nb}\approx 8.0 - 8.5$ K), show that the films are 
indeed of good quality.

At voltages below the gap of Nb, a series of gate-dependent resonance peaks appears in the $IV$ 
characteristics 
of the SET (Fig.~\ref{fig1}). We interpret this as resonant tunneling of Cooper pairs, a transport 
phenomenon first predicted theoretically and later observed in Al symmetrically-biased 
superconducting SETs.\cite{haviland} Below we describe the same process for our Nb-island SETs under 
the asymmetric bias shown in Fig.~\ref{fig1}. We consider a generic process in which a charge $\delta 
q_{1}$ tunnels through the left junction and a charge 
$\delta q_{2}$ tunnels through the second junction, both into the island. During the process, 
a charge $\delta q = \delta q_{1} + \delta q_{2}$ is transferred into the island and a charge $\delta 
Q = \delta q_{1} - \delta q_{2}$
is transferred through the external circuit in the forward direction.
The change in the electrostatic free energy (including work done by the sources)
associated with this process 
is 
\begin{eqnarray}
E(\delta q,\delta Q) &=& \frac{(\delta q)^2}{2C_{\Sigma}} + \frac{\delta 
q}{C_{\Sigma}}(q_{0}-C_\mathrm{g}V_\mathrm{g}) \nonumber \\ &&
+\frac{\delta q}{C_{\Sigma}}\left(C_\mathrm{g}-\frac{C_{2}-C_{1}}{2}\right)V - \frac{\delta Q}{2}V,
\end{eqnarray}
where $q_{0}$ is the initial charge of the island, $C_1$ and $C_2$ are the capacitances of the left 
and right junctions, $C_\mathrm{g}$ is the gate capacitance, and 
$C_{\Sigma}=C_{1}+C_{2}+C_\mathrm{g}$. 
Resonant Cooper pair tunneling in superconducting SETs occurs when no energy is required for 
processes resulting in the transport of $m$ Cooper pairs $\delta Q = -2me$ through the external 
circuit and the creation of an excess of $n$ Cooper pairs $\delta q = -2ne$ on the island, 
i.e.,~$E(-2ne,-2me) = 0$. 

The dominant processes 
are those involving $m=\pm 1$ and $n=\pm 1$, corresponding to a single Cooper pair tunneling through 
either one of the junctions.\cite{haviland} 
For these processes, the resonant condition describes two lattices with a cell size of 
$(4e/C_{\Sigma})\!\times\!(2e/C_\mathrm{g})$ (in $V\! \times\!V_\mathrm{g}$ plane) corresponding to 
odd and even values of $q_{0}/e$, and displaced with respect to each other by $2e/C_{\Sigma}$ along 
the $V$ axis and by $e/C_\mathrm{g}$
along the the $V_\mathrm{g}$ axis. During the time of an experiment, $q_0$ can either be fixed, in 
which case 
only one lattice pattern will appear, or it can
fluctuate between odd and even values, in which case what will
be measured is the overlap between the two lattices, resulting in a checkerboard lattice with half 
the periodicity.
This last situation occurs indeed in all of our three samples. To check this, we first measured the 
gate modulation at voltages above the quasiparticle threshold of Nb, where the transport is of 
single-electron type, and determined $e/C_\mathrm{g}$. This corresponds to the size of the 
checkerboard pattern along $V_\mathrm{g}$ that we see in our measurements at lower
bias as well, e.g., $e/C_\mathrm{g} = 12.3$ mV ($C_ \mathrm{g}=12.9$ aF) for the sample of 
Fig.~\ref{fig1}. (For clarity, we will concentrate our discussion on that particular sample from now 
on. The other two samples yielded similar results.) The observation that the checkerboard pattern is 
$1e$-periodic allows us to determine $e/C_{\Sigma}=68.3$ $\mu$V  and thus the charging energy 
$E_\mathrm{C}=e^{2}/2C_{\Sigma}=34.1$ $\mu$eV.\cite{joyez} 
The SET's junction asymmetry is reflected in the different absolute values of the positive-slope 
lines (for which $n$ and $m$ have different signs) versus the negative-slope lines ($m$ and $n$ have 
the same sign), yielding $C_{1}=1.03$ fF and $C_{2}=1.29$ fF.

\begin{widetext}
\hspace{50mm}
\begin{figure}[htb]
\includegraphics[width=140truemm]{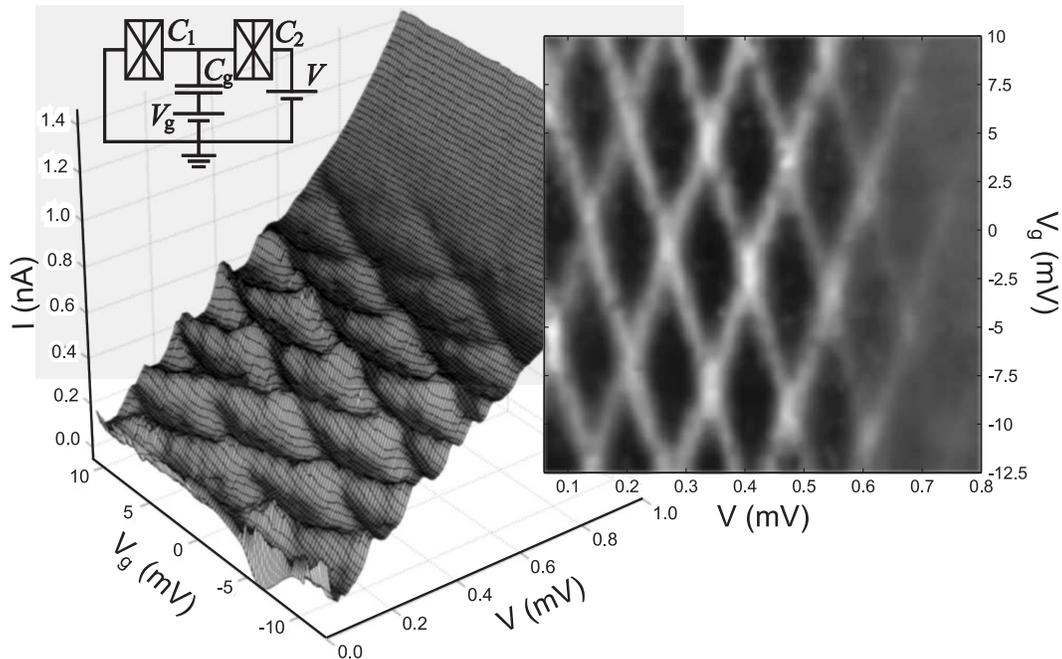}
\caption{\sl Schematic of the circuit (upper left inset) and current versus bias and gate voltages 
(3D plot), showing the checkerboard pattern (right inset) associated with resonant tunneling of 
Cooper pairs. The contour plot of the right inset is obtained by subtracting the background current 
thus leaving only the characteristics of the resonant tunneling.}
\label{fig1}
\end{figure} 
\end{widetext}

To check for consistency, we have determined $E_\mathrm{C}$ also by standard Coulomb-blockade 
thermometry \cite{cbt} with the superconductivity both in  Al and Nb suppressed. We find
$E_\mathrm{C}=36$ $\mu$eV, which is in very close agreement with the results above.

\begin{figure}[th]
\includegraphics[width=77truemm]{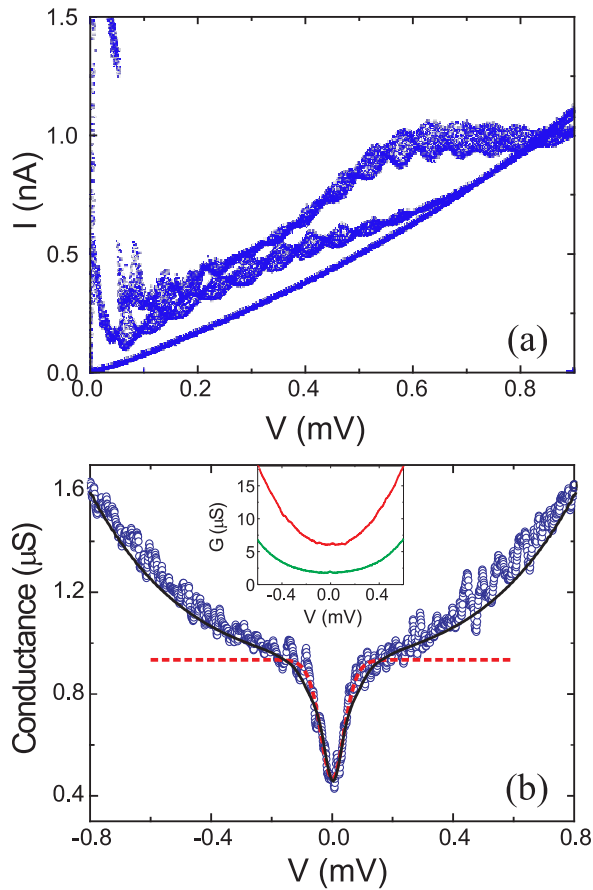}
\caption{\sl (a) Effect of a magnetic field on $IV$ characteristics (current values at different 
$V_\mathrm{g}$'s, showing gate modulation, are superposed): zero, upper curve; intermediate value, 
middle curve; complete suppression of the gap of Al, lower curve. (b) The main graph shows the 
low-bias Coulomb blockade dip: experimental values (blue), standard BCS Coulomb-blockade thermometry 
fit (red dotted line), and the predictions of the orthodox theory with lifetime broadening and gap 
inhomogeneity (black continuous line).
The upper inset shows the conductances for two single Al-Nb junctions of 
normal resistances 30 k$\Omega$ (lower curve) and 11 k$\Omega$ (upper curve).}
\label{fig2}
\end{figure}

The second intriguing feature in the low-bias $IV$ of our Nb-island SETs, is the presence of 
relatively large subgap currents: from figures \ref{fig1} and \ref{fig2}(a) one can see that the 
Cooper pair resonances appear as being built upon a current which increases with bias voltage (at 
voltages larger than $2\Delta_{\rm Al} + 2\Delta_{\rm Nb}$, it will merge into the quasiparticle 
current).
The origin of the subgap currents in superconductors is still a matter of intense theoretical and 
experimental investigations.\cite{review}
Here, we further explore these currents by using a magnetic field to suppress the gap of Al; the 
transition to the normal-metal state in the leads is clearly indicated by the disappearance of the 
Josephson effect and of the Cooper pair resonances as seen in Fig.~\ref{fig2}(a). 

While Al is in the normal state, a very interesting feature appears at low bias voltages 
[Fig.~\ref{fig2}(b)], where we observe a dip in the conductance. In measurements on single Al-Nb 
junctions [again with the gap of Al suppressed, Fig.~\ref{fig2}(b) inset] this feature is not 
present.\cite{us} Instead, below 0.2 mV
the $IV$ is approximately ohmic. This shows that the dip is due to the Coulomb blockade of the subgap 
current, and that in the first approximation one could attempt to fit the graph with the standard 
analytical expressions for Coulomb-blockade thermometry \cite{cbt} (CBT) [Fig.~\ref{fig2}(b), main 
graph, red dashed line]. Fitting yields $T \approx 98$ mK, which agrees well with the value measured 
using a calibrated resistor thermally anchored to the mixing chamber, and $E_\mathrm{C}\approx 13$ 
$\mu$eV, which is in reasonable agreement with the value determined from Cooper pair resonances, 
considering that we are not in the limit $k_\mathrm{B}T\gg E_\mathrm{C}$ and therefore the standard 
approximations of Coulomb blockade thermometry are not accurate, especially for determining 
$E_\mathrm{C}$.\cite{cbt}

A better model should take into account that we are at low temperatures $k_\mathrm{B}T\!<\! 
E_\mathrm{C}$ and also describe the nonlinear increase in the subgap current at higher voltages.
To develop such a model, we introduce a lifetime broadening $\Gamma$ of the quasiparticle energies, 
resulting in a density of states \cite{mi}
\begin{equation}
\rho (E, \Delta_\mathrm{Nb}) = \left\vert {\rm Re} \left( \frac{E - i \Gamma}{\sqrt{(E- i \Gamma )^2 
- \Delta_{\rm Nb}^2}} \right) \right\vert .
\end{equation} 
This effect could be caused by the proximity to the metal-insulator transition, due to the granular 
structure of the Nb films (with our films being still on the metalic side, $\Gamma\!\ll 
\!\Delta_\mathrm{Nb}$).\cite{mi}
Another possibility is the opening of conduction channels in the junctions,\cite{joyez} as suggested 
by the relatively large value of the room-temperature conductance [for the sample presented in 
Fig.~\ref{fig1}, Fig.~\ref{fig2}, and Fig.~3(a), this was 30 $\mu$S].
Irrespective to the microscopic origin,  
this density of states accounts well for the existence of subgap currents seen in 
Fig.~\ref{fig2}(b),inset. At low energies $E \!\ll\! \Delta_\mathrm{Nb}$ the density of states can be 
approximated as $\rho (E,\Delta_{Nb}) \approx 
\Gamma /\Delta_{\rm Nb} + (3\Gamma /2\Delta )(E/\Delta_{\rm Nb})^2$. In the case of a single junction 
of resistance $R$ with a normal-state metal this results in a subgap conductance $R^{-1}\rho 
(V,\Delta_\mathrm{Nb})$; therefore, ohmic 
for low enough voltages and increasing as $V^2$ for relatively larger voltages. In addition, the 
granular structure of the films and, possibly, impurities resulting from outgassing of the mask 
polymer is also making the film inhomogenous. As a result, the gap edge is smeared due to local 
fluctuations in the effective electron-electron interaction. Other fabrication techniques [e.g. the 
use of stronger polymers such as PES \cite{zorin}] could improve the quality of the Nb films. 
For our samples, since the superconducting coherence length in bulk Nb is only
$\xi_\mathrm{Nb} = 38$ nm, we expect to be in the limit in which the size of the inhomogeneities is 
much larger than $\xi_\mathrm{Nb}$. In this case, it has been shown \cite{smeared} that the effective 
(average) density of states takes the form
\begin{equation}
\bar{\rho}
(E) = \int_{0}^{\infty} d\Delta_\mathrm{Nb} {\cal P}(\Delta_\mathrm{Nb})\rho (E,\Delta_\mathrm{Nb}),
\end{equation}
where ${\cal P}$ is the probability density associated with a certain value of the gap. For small 
levels of inhomogeneity, the function ${\cal P}$ is a Gaussian of standard distribution $\sigma$ 
centered around an average
gap value (slightly smaller than the bulk value). 

Finally, to complete our model, we calculate numerically the  currents and conductances for the 
superconducting SET in the framework of the orthodox theory of sequential tunneling, by 
computing the tunneling probabilities into and out of the island (with the density of states above 
for Nb) and by solving the master equation for the charge on the island. In the case of suppressed Al 
gap, a very good fit is obtained for $\sigma = 0.38$ meV, $\Gamma = 37.5$ $\mu$eV, $T = 95.7$ mK, and 
charging energy, $E_\mathrm{C} = 34.1$ $\mu$eV, as determined before [Fig.~\ref{fig2}(b), main graph, 
black continuous line]. The temperature also agrees well with the value obtained by simple CBT 
fitting above and the value of the thermometer.

\begin{figure}[htb]
\includegraphics[width=87truemm]{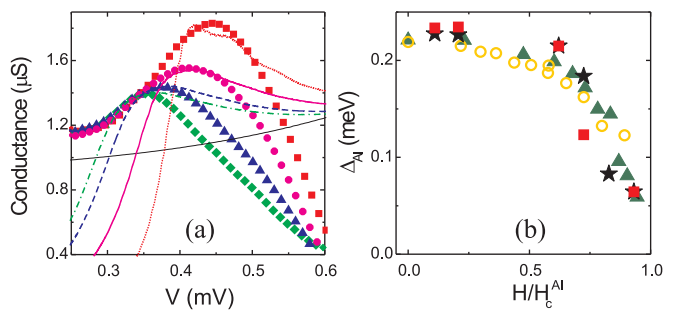}
\caption{\sl (a) Qualitative comparison between the conductances extracted from the data by averaging 
out the gate modulation
(symbols) and those obtained by numerical simulations (lines) at different magnetic fields. (b) The 
gap of Al leads (open circle), measured separately, plotted together with the Al gaps determined from 
the conductance peaks of (a) for three SET samples (square, triangle, and star).}
\label{fig3}
\end{figure}  

In the case when the leads are superconducting the comparison with the experimental data can be done 
only qualitatively: the main reason is the existence of Cooper pair resonances peaks, yielding an 
extra contribution to the current which cannot be eliminated in a straightforward way. However, a 
number of qualitative features can still be observed by averaging the resonance peaks over gate 
voltages. The conductances obtained by this procedure should reflect the voltage dependence of the 
subgap current in the region where the height of the resonance peaks is approximately constant. A 
reasonably fair
qualitative agreement with the model presented above is obtained [Fig.~\ref{fig3}(a)].
The main feature that we see both in the calculated conductivities and in the 
conductivities obtained from our data through the above procedure is the appearance of a peak at the 
onset of the Al 
quasiparticle threshold ($2\Delta_{\rm Al}$). By using a magnetic field to partially suppress the gap 
of Al, we see that the 
peak moves to the left [see Fig.~\ref{fig2}(a) and Fig.~\ref{fig3}(a)]. To verify that this is indeed 
the case, we 
have measured single Al-Al junctions with the same fabrication parameters as the leads of the 
Nb-island SET, and determined the gap of Al at various magnetic fields [Fig.~\ref{fig3}(b), circles]. 
The agreement with the gaps determined from 
the low-voltage features [Fig.~\ref{fig3}(a)] of three Nb-island SET samples (square, triangle, star) 
is good.

In conclusion, we have fabricated and measured superconducting single-electron transistors and we 
have presented evidence for a number of transport processes occurring at bias voltages below the gap 
of the island: resonant tunneling of Cooper pairs, Coulomb blockade of the subgap current for normal 
leads, and the appearance of a step
in the subgap current at $2\Delta_\mathrm{Al}$ for the case of superconducting leads.

\acknowledgments

This work was supported by the Academy of Finland (Acad. Res. Grant no. 00857 and Projects No. 
7111994, 7118122, 7205476), EU (IST-1999-10673, HPMF-CT-2002-01893), and the Finnish National 
Graduate School in Materials Physics. J.K. acknowledges the support of the Department of Energy, 
Office of Basic 
Energy Sciences via the Chemical Sciences, Geosciences, and Biosciences 
Division. G.~S.~P. would like to thank D.~V.~Averin, Yu.~V.~Nazarov, and I.~S.~ Beloborodov for 
useful discussions. 

\vspace{-5mm}
\footnotesize
\bibliography{pump}

\end{document}